\documentclass[conference,a4paper]{IEEEtran}
\IEEEoverridecommandlockouts

\usepackage{xr-hyper} 
\usepackage{hyperref} 
\usepackage[acronym, nonumberlist, nogroupskip, style=super]{glossaries}
\usepackage{hyperref}
\newacronym{iot}{IoT}{Internet of Things} 

\usepackage{cite}
\usepackage{amsmath, amssymb, amsfonts}
\usepackage{graphicx}
\usepackage{textcomp}
\usepackage{bmpsize}
\usepackage[dvipsnames]{xcolor}
\usepackage{lipsum}

\usepackage{multirow}
\usepackage{tabularx}

\usepackage[T1]{fontenc}

\usepackage{subfig}	
\usepackage{soul}
\usepackage{booktabs}

\usepackage{multirow}
\usepackage{rotating}
\usepackage{booktabs}

\usepackage{color, colortbl}

\usepackage{comment}

\usepackage{cite}
\usepackage{amsmath, amssymb, amsfonts}
\usepackage{algorithmic}
\usepackage{graphicx}
\usepackage{textcomp}
\usepackage{xcolor}

\def\BibTeX{{\rm B\kern-. 05em{\sc i\kern-. 025em b}\kern-. 08em
 T\kern-. 1667em\lower. 7ex\hbox{E}\kern-. 125emX}}
\begin{document}

\title{Graph-Based Floor Separation Using Node Embeddings and Clustering of WiFi Trajectories\
	\thanks{}
}

\author{
    \IEEEauthorblockN{ Rabia Yasa Kostas\IEEEauthorrefmark{1}, Kahraman Kostas\IEEEauthorrefmark{2}}
      \IEEEauthorblockA{\IEEEauthorrefmark{1} Department of Computer Technologies, Gumushane University ,Gumushane, Turkiye}
    \IEEEauthorblockA{\IEEEauthorrefmark{2}School of Mathematical \& Computer Sciences, Heriot Watt University, Edinburgh, United Kingdom}

	{rabia.yasakostas@gumushane.edu.tr, kk97@hw.ac.uk}
}

\maketitle
\makeatletter
\def\ps@IEEEtitlepagestyle{%
  \def\@oddfoot{\mycopyrightnotice}%
  \def\@oddhead{\hbox{}\@IEEEheaderstyle\leftmark\hfil\thepage}\relax
  \def\@evenhead{\@IEEEheaderstyle\thepage\hfil\leftmark\hbox{}}\relax
  \def\@evenfoot{}%
}
\def\mycopyrightnotice{%
}
\makeatother
\maketitle
\begin{abstract}
Indoor positioning systems (IPSs) are increasingly vital for location-based services in complex multi-storey environments. This study proposes a novel graph-based approach for floor separation using Wi-Fi fingerprint trajectories, addressing the challenge of vertical localization in indoor settings. We construct a graph where nodes represent Wi-Fi fingerprints, and edges are weighted by signal similarity and contextual transitions. Node2Vec is employed to generate low-dimensional embeddings, which are subsequently clustered using K-means to identify distinct floors. Evaluated on the Huawei University Challenge 2021 dataset, our method outperforms traditional community detection algorithms, achieving an accuracy of 68.97\%, an F1-score of 61.99\%, and an Adjusted Rand Index of 57.19\%. By publicly releasing the preprocessed dataset and implementation code, this work contributes to advancing research in indoor positioning. The proposed approach demonstrates robustness to signal noise and architectural complexities, offering a scalable solution for floor-level localization.

\end{abstract}


\section{Introduction}

Indoor positioning has garnered significant attention in recent years, driven by rapid technological advances and the growing reliance on indoor location-based services. As urbanization accelerates, a substantial portion of human activity now occurs within indoor environments such as shopping malls, airports, offices, and hospitals~\cite{Zhao2024Research}. This trend has led to an increased demand for reliable and accurate indoor positioning systems (IPSs), which are essential for applications including indoor navigation, security management, asset tracking, and personalized customer services. Unlike outdoor systems that rely on Global Positioning System (GPS) signals, IPSs face unique challenges due to signal obstructions caused by walls, furniture, and complex architectural layouts~\cite{Geok2020}.

A particularly complex issue in indoor positioning is vertical localization—determining a user's floor level in multi-storey buildings. Signal leakage between floors, especially with Wi-Fi and other radio signals, can lead to overlapping signal patterns, making it difficult to accurately identify the correct floor. This issue is further compounded in open architectural spaces such as atriums and balconies, where signals reflect and propagate across multiple levels~\cite{Alitalesi2022}. Environmental dynamics, user movement, and interference from nearby devices introduce additional variability, further complicating vertical positioning~\cite{Rizk2020}.

Among various IPS technologies, Wi-Fi fingerprinting has emerged as a prominent solution due to its low deployment cost and ability to utilize existing wireless infrastructure. This technique typically involves two phases: an offline phase, where signal strength measurements from access points are collected to build a fingerprint database, and an online phase, where real-time signals are matched against this database to estimate the user's location~\cite{Ali19wi}. Despite its advantages, Wi-Fi fingerprinting is susceptible to signal fluctuation and multipath interference, which can reduce localization accuracy. To address these limitations, recent studies have incorporated machine learning models that improve robustness to signal noise and variability~\cite{Tinh21Ensemble, li19Nrssd}.

In parallel, graph-based approaches are gaining traction in indoor localization due to their ability to model complex spatial and signal relationships. In these methods, Wi-Fi fingerprints are represented as nodes in a graph, while the relationships—based on signal similarity or contextual transitions like stairways and elevators—are captured by weighted edges~\cite{kim23Gconvloc}. This graph structure allows for a more nuanced representation of the indoor environment compared to traditional techniques.

In this study, we introduce a novel graph-based approach for clustering building floors using Wi-Fi fingerprint similarity. We construct a graph where each node corresponds to a Wi-Fi fingerprint, and edges are weighted based on signal distance metrics and contextual information. To extract meaningful representations from this graph, we apply Node2Vec~\cite{grover2016node2vec}, a graph embedding technique that learns low-dimensional vector embeddings of nodes. These embeddings are then clustered to distinguish between different floor levels. Our method is evaluated using the Huawei University Challenge 2021 dataset—a comprehensive yet underexplored dataset previously available only to competition participants. As part of this work, we preprocess and publicly release the dataset, along with our implementation code, to support further research in the community\footnote{Source code and dataset available at:\href{https://github.com/kahramankostas/IPIN2025}{github.com/kahramankostas/IPIN2025}}. 

The remainder of this paper is organized as follows: Section 2 reviews related work, Section 3 details the proposed methodology, Section 4 presents experimental results and analysis, and Section 5 concludes with a discussion and future directions.

\section{Related Work} \label{Related Work}

Floor detection in indoor environments is a long-standing challenge, particularly in multi-story buildings where vertical localization is critical. Various sensor-based and data-driven approaches have been explored in the literature, including barometric pressure sensing, Inertial Measurement Units (IMUs), WiFi Received Signal Strength Indicator (RSSI) data, and machine learning techniques. This section reviews prominent approaches and highlights the relative advantages of WiFi RSSI-based clustering, which forms the foundation of our proposed method.

\subsection{Barometric Pressure Sensors}
Barometric pressure sensors are frequently used for altitude estimation and have been integrated into floor detection systems due to the inverse relationship between atmospheric pressure and elevation. As users move between floors, the ambient pressure changes, providing a basis for vertical localization. Huang et al. demonstrated that combining barometric data with WiFi signals and motion detection from accelerometers can achieve floor detection accuracies of up to 99.2\%~\cite{Huang20}. Similarly, the "pressure-pair" technique proposed by Yi et al. leverages pressure variations in conjunction with smartphone sensors to improve localization accuracy~\cite{Yi19}. However, such systems often require fixed reference sensors deployed at known altitudes to correct for environmental fluctuations, which can limit scalability and increase infrastructure costs~\cite{10.1109/tim.2018.2863978}.

\subsection{IMU-Based Elevation and Step Detection}
Inertial Measurement Units (IMUs), which integrate accelerometer and gyroscope data, contribute to vertical positioning by detecting patterns in user movement, such as step frequency and vertical displacement. These signals offer a complementary layer to barometric systems. Studies have shown that integrating IMU and barometric data can significantly improve reliability and reduce sensitivity to noise~\cite{10.1109/jsen.2018.2852494, Fetzer18}. The fusion of these two modalities can effectively mitigate external disturbances such as pressure drift, though IMU-based methods may struggle with stationary users and require calibration for individual movement profiles~\cite{10.1109/tim.2018.2863978}.

\subsection{WiFi RSSI-Based Floor Detection}
WiFi RSSI data is increasingly used for indoor floor estimation, particularly in urban environments where dense wireless infrastructure is available. Fingerprinting-based approaches utilize variations in signal strength across locations to identify specific floors. Recent advances in deep learning have improved the ability to learn meaningful representations from RSSI patterns and have led to enhanced performance in floor classification tasks~\cite{Zhou24, Ayinla24}. These models are robust to traditional signal degradation phenomena such as multipath fading and shadowing.

Importantly, methods based on WiFi distance estimation and clustering present significant advantages over conventional techniques. 
WiFi distance estimation refers to the process of determining the distance between a WiFi-enabled device and one or more WiFi access points based on various signal characteristics, such as received signal strength (RSS)~\cite{kostas2022wifi}. By clustering RSSI-based distance estimates, spatially coherent groupings can be formed that reduce the impact of sensor noise and user-specific movement variability. Unlike barometric and IMU-based systems, these approaches rely solely on existing wireless infrastructure, making them highly scalable and hardware-independent. In this context, the clustering strategy employed in this study capitalizes on the ubiquitous nature of WiFi and leverages spatial signal consistency to provide a more practical and robust solution for floor identification.

\subsection{Machine Learning-Based Classification Techniques}
Recent research has demonstrated that machine learning algorithms can significantly enhance floor detection accuracy by processing multimodal sensor data. Classifiers trained on data from barometers, IMUs, and WiFi networks are capable of learning complex spatial patterns and distinguishing floors more effectively~\cite{Shi20}. Convolutional Neural Networks (CNNs), for instance, have been successfully applied to time-series RSSI data, mitigating spatial ambiguity and noise effects~\cite{Liu21}. Integrating these models with traditional sensing methods yields hybrid systems capable of achieving high performance even in architecturally complex indoor environments~\cite{Zhou24, Li24}.

Graph-based methods have emerged as powerful tools in indoor positioning, particularly for structuring and interpreting complex spatial environments. By representing signal-based relationships—such as WiFi RSSI-derived distance estimates—as weighted graphs, these methods facilitate efficient partitioning of spatial data using clustering algorithms. Community detection techniques like Fast-Greedy, Infomap, Label Propagation, Louvain, and Leiden are widely used in graph partitioning tasks. Among these, the Louvain algorithm is notable for its computational efficiency and has been successfully applied to network structures such as brain connectivity data~\cite{10.1016/j.physrep.2009.11.002, 10.1089/brain.2018.0603}. The Infomap algorithm also shows strong performance in identifying hierarchical structures within networks~\cite{10.1089/brain.2018.0603}.

In the context of indoor localization, clustering algorithms have been employed to enhance positioning accuracy in GPS-denied environments~\cite{10.1049/iet-com.2017.1055}. By detecting community structures within signal graphs, these methods can model spatial partitions corresponding to different physical floors or rooms. Studies have demonstrated that such clustering strategies improve spatial reasoning and navigation planning~\cite{10.1080/13658816.2015.1041141}. Comparative analyses by Anuar et al.\cite{10.1088/1742-6596/2129/1/012028} and Li et al.\cite{10.2991/ijcis.d.210329.001} suggest that while both Louvain and Leiden algorithms offer strong baseline performance, the Leiden method provides enhanced scalability and community quality, making it more suitable for large-scale indoor positioning tasks.

Complementing clustering approaches, graph embedding techniques such as Node2Vec, DeepWalk, and LINE have demonstrated their capacity to capture latent structural features in graphs. These embeddings generate vector representations of nodes that preserve both local and global graph topology. Node2Vec, in particular, balances exploration and exploitation in graph walks and has proven effective in various domains including community detection and structural similarity analysis~\cite{10.1016/j.physa.2015.09.093, 10.1007/978-3-319-55471-6_1}. Recent adaptations of Node2Vec for indoor localization have shown promise in floor detection, especially when spatial relationships between WiFi reference points are embedded as proximity graphs.

Despite their proven effectiveness, graph embedding strategies remain underutilized in the specific context of WiFi-based floor estimation. Existing literature suggests a research gap in evaluating these embeddings for indoor applications, particularly in relation to graph clustering techniques that leverage RSSI-derived distance graphs. Addressing this gap, our proposed method integrates distance-based graph construction with community detection algorithms, offering a scalable and hardware-independent approach to floor identification. As highlighted by recent work on signed graph embeddings and non-linear error modeling~\cite{10.1145/3539597.3570401, 10.3390/jsan8020021}, further investigation into task-specific embedding optimizations could lead to notable gains in accuracy and robustness for floor-level localization.

\section{Methodology}
In this section, dataset, feature extraction, features, ML algorithms and evaluation metrics are explained. 
\subsection{Problem Definition}

The goal of this study is to assign each WiFi fingerprint trajectory to its correct floor in a multi-storey indoor environment using a graph-based clustering approach. The problem setting is inspired by the Huawei University Challenge 2021 Task 2, and it involves the following inputs:

\begin{itemize}
	\item A set of WiFi fingerprint observations collected from various indoor locations within a shopping mall.
	\item Step connections (i.e., sequential transitions) between consecutive fingerprints belonging to the same trajectory.
	\item Elevation hints (e.g., elevator or staircase usage) that indicate certain fingerprint pairs are on different floors.
	\item Precomputed, though potentially noisy, pairwise WiFi distance estimations derived from signal strength.
\end{itemize}

The objective is to cluster trajectories such that all fingerprints within a cluster belong to the same floor, without prior knowledge of the number of floors. This makes the task a form of unsupervised learning with weak supervision. The number of floors varies across buildings and is unknown during inference, ranging between 3 and 20 floors as per competition guidelines.

One of the core challenges lies in signal ambiguity due to architectural features like balconies or atriums, where fingerprints on different floors may exhibit similar signal characteristics. Moreover, traditional fingerprint-based classifiers struggle to generalize due to signal variability and the presence of spatially similar but vertically distinct points.

\subsection{Huawei University Challenge 2021 Dataset}

The experiments in this study are conducted using the publicly available dataset provided in the Huawei University Challenge 2021. The dataset is designed for floor prediction in large-scale indoor environments and consists of multiple files providing structural and signal-level information:

\begin{itemize}
	\item \texttt{fingerprints.json}: Contains individual WiFi fingerprints with associated access point signal strengths and locations.
	\item \texttt{steps.csv}: Lists sequential step pairs (i.e., motion edges) within the same trajectory, assumed to be on the same floor.
	\item \texttt{elevations.csv}: Contains fingerprint pairs that involve vertical transitions (e.g., elevators or stairs), confirming these are from different floors.
	\item \texttt{estimated\_wifi\_distances.csv}: Includes noisy, pre-estimated distances between WiFi fingerprints based on signal strength measurements.
	\item \texttt{lookup.json}: Provides a mapping between fingerprint IDs and their corresponding trajectory IDs.
	\item \texttt{GT.json}: Supplies ground-truth floor labels for a subset of trajectories, useful for training and validation.
\end{itemize}

Importantly, the training and test sets originate from different buildings, which introduces domain shift and necessitates generalization across distinct structural layouts. Furthermore, the output of the task is not per-fingerprint classification, but rather the correct clustering of entire trajectories into floor-based groups. The dataset is clean in the sense that it contains no outlier fingerprints, and all given trajectories are valid.

\subsection{Clustering and Community Detection Methods}

In this study, we employ several machine learning techniques with a primary focus on community detection and clustering in complex networks. Each method offers unique advantages depending on the underlying data structure and desired granularity of the partitioning. The selected algorithms include Louvain, Leiden, Label Propagation, Infomap, Fast Greedy, K-means, and Node2Vec.

\subsubsection*{Louvain Method}
The Louvain algorithm is a widely used method for community detection that operates by optimizing modularity in a two-phase process. Initially, each node is assigned to its own community, and local modularity gains are used to iteratively merge communities. In the second phase, the detected communities are aggregated into super-nodes, and the process repeats until convergence. This method is particularly efficient for large networks and is recognized for producing high-quality partitions \cite{10.1038/srep02216,10.2991/ijcis.d.210329.001}.

\subsubsection*{Leiden Method}
An enhancement over Louvain, the Leiden algorithm improves both the quality and robustness of detected communities. It addresses known limitations in the Louvain approach, such as disconnected communities and resolution limits. Leiden ensures well-connected communities by incorporating refinement phases and better initialization heuristics \cite{10.1038/srep02216,10.1088/1742-6596/2129/1/012028}. As a result, it is particularly effective in applications requiring high-resolution community structure detection.

\subsubsection*{Label Propagation}
Label Propagation Algorithm (LPA) detects communities by propagating labels through the network in an iterative fashion. At each iteration, nodes update their label based on the majority label among their neighbors. LPA is parameter-free and computationally efficient, making it suitable for large-scale networks \cite{10.1371/journal.pone.0051947,10.18280/i2m.180202}. Recent improvements and parallel implementations of LPA have significantly enhanced its performance and applicability to dynamic and overlapping communities \cite{10.48550/arxiv.2301.09125}.

\subsubsection*{Infomap}
Infomap employs principles of information theory and random walks to reveal community structures. By minimizing the expected description length of a random walker's path through the network, it identifies modules that correspond to real information flow. Its capacity to detect hierarchical and fine-grained structures is especially valuable in large and complex networks \cite{10.1155/2015/934301,10.1109/icdmw.2013.138}.

\subsubsection*{Fast Greedy}
Fast Greedy is a modularity-based method that iteratively merges communities in a greedy fashion to optimize global modularity. Although less fine-tuned than Leiden or Infomap, it offers a favorable trade-off between speed and accuracy, making it useful for exploratory analysis on large datasets \cite{10.1038/srep02216,10.1016/j.physa.2015.09.093}.

\subsubsection*{K-Means}
K-means is a centroid-based clustering algorithm that partitions data points into \(k\) clusters by minimizing intra-cluster variance. While not inherently designed for graph data, it can be applied when nodes are embedded into a vector space, such as via Node2Vec. K-means is most effective when the number of clusters is known or can be estimated in advance \cite{10.1155/2022/2749091}.

\subsubsection*{Node2Vec}
Node2Vec generates low-dimensional embeddings for nodes in a network by simulating biased random walks. These embeddings preserve network topology and are suitable for various downstream tasks such as node classification and clustering. The flexibility in exploring homophilic and structural equivalence makes Node2Vec a powerful tool for learning network representations \cite{10.25236/ajcis.2022.051212}.

\subsubsection*{Summary}
Together, these methods provide a diverse toolbox for community detection and clustering in network-structured data. The choice of method depends on factors such as the scale of the network, the expected number of clusters, and the desired resolution of the community structure \cite{10.1016/j.physrep.2009.11.002,10.1089/brain.2018.0603}.

\subsection{Evaluation Metrics}

In order to assess the performance of the clustering algorithms, several evaluation metrics were employed. These metrics provide insights into the quality of the clustering results by comparing the predicted clusters with the ground truth labels. The following metrics were used:

\subsubsection{Accuracy}
Accuracy measures the proportion of correctly assigned instances compared to the total number of instances. Specifically, it evaluates how well the predicted clusters match the actual clusters. A higher accuracy indicates a closer alignment with the ground truth.

\[
\text{Accuracy} = \frac{\text{Number of correct predictions}}{\text{Total number of instances}}
\]

\subsubsection{F1-Score}
The F1-score is the harmonic mean of precision and recall, which are two key metrics in classification tasks. Precision quantifies how many of the predicted positive instances are actually positive, while recall measures how many of the actual positive instances were correctly identified. The F1-score balances both precision and recall into a single metric, making it especially useful when there is an uneven class distribution.

\[
\text{F1-Score} = 2 \cdot \frac{\text{Precision} \cdot \text{Recall}}{\text{Precision} + \text{Recall}}
\]

\subsubsection{Adjusted Rand Index (ARI)}
The Adjusted Rand Index (ARI) measures the similarity between the predicted clustering and the ground truth clustering, adjusted for chance. The ARI ranges from -1 to 1, where 1 indicates perfect agreement, 0 indicates random clustering, and negative values indicate worse than random clustering. The ARI is particularly useful in evaluating the quality of the clustering results in comparison to a ground truth, accounting for random chance.

\[
\text{ARI} = \frac{\text{RI} - \mathbb{E}[\text{RI}]}{\max(\text{RI}) - \mathbb{E}[\text{RI}]}
\]
Where \(\text{RI}\) is the Rand Index, and \(\mathbb{E}[\text{RI}]\) is the expected Rand Index under a random distribution.

\subsubsection{Normalized Mutual Information (NMI)}
Normalized Mutual Information (NMI) quantifies the amount of shared information between the predicted clusters and the ground truth clusters. It measures the dependence between the two clustering solutions, with a higher NMI indicating a greater degree of shared information. The NMI value ranges from 0 (no shared information) to 1 (perfect agreement).

\[
\text{NMI} = \frac{I(\text{Predicted}, \text{True})}{\sqrt{H(\text{Predicted}) H(\text{True})}}
\]
Where \(I\) denotes mutual information, and \(H\) represents entropy.

\subsubsection{Average Cluster Purity}
Cluster purity measures the extent to which each cluster contains instances from a single ground truth class. The average cluster purity is calculated by averaging the purity of each cluster. A higher purity indicates that each cluster predominantly contains instances from a single class, suggesting better clustering quality. However, high purity alone does not guarantee that the clustering captures the true structure of the data.

\[
\text{Purity} = \frac{1}{N} \sum_{i=1}^{k} \max_j |C_i \cap T_j|
\]
Where \(N\) is the total number of instances, \(k\) is the number of clusters, \(C_i\) is the set of instances in cluster \(i\), and \(T_j\) is the set of instances from the true class \(j\).

These evaluation metrics provide a comprehensive assessment of the clustering performance, highlighting the strengths and weaknesses of each algorithm in relation to the ground truth.
\section{Experiments}

The experimental process began by converting the input files into a graph structure, where nodes represent individual trajectories or Access Points (APs), and edge weights encode their estimated proximity—incorporating both horizontal distance and elevation information to reflect real-world spatial adjacency. This graph serves as the shared input for all evaluated methods.

\begin{figure}[ht]
	\centerline{\includegraphics[width=1\columnwidth]{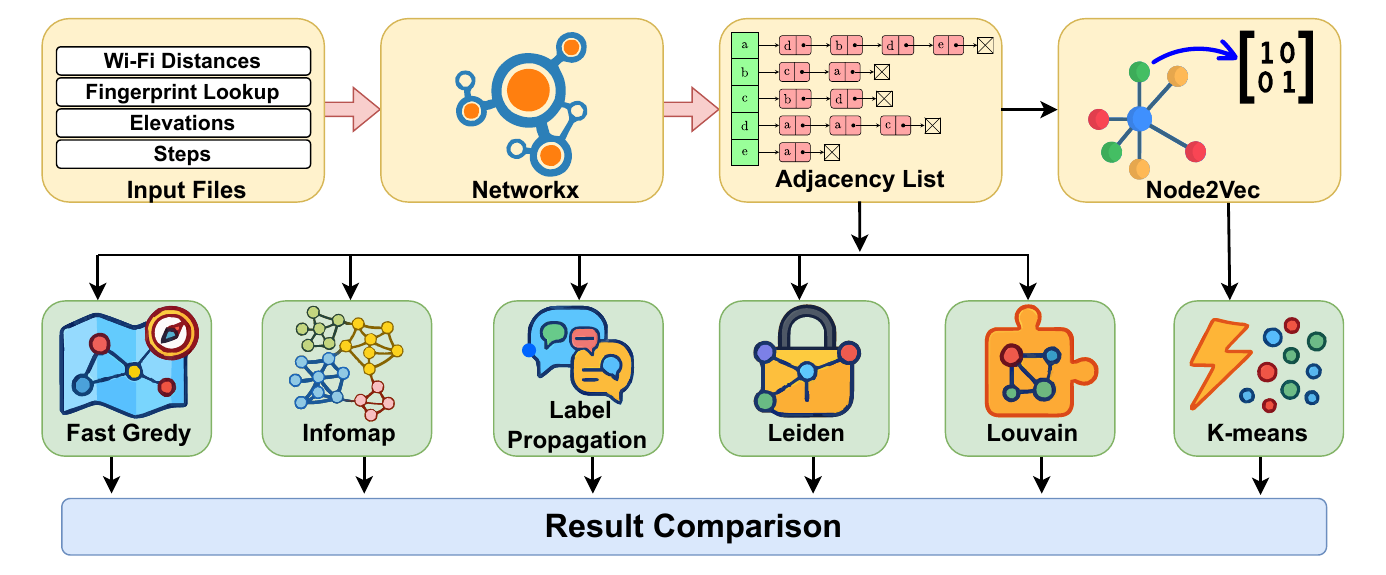}}
	\caption{Overview of the experimental workflow. The trajectory graph is used as input for both the baseline community detection algorithms (Louvain, Leiden, Infomap, Label Propagation, Fast Greedy) and the proposed Node2Vec+KMeans pipeline. While baseline methods operate directly on the graph structure, the proposed approach first transforms the graph into a 32-dimensional embedding space using Node2Vec, followed by clustering with K-Means.}
	\label{fig:workflow}
\end{figure}

As illustrated in Figure~\ref{fig:workflow}, five baseline community detection algorithms—Louvain, Leiden, Label Propagation, Infomap, and Fast Greedy—were applied directly to the trajectory graph. These methods attempt to detect densely connected subgraphs, relying solely on topological properties such as modularity or label diffusion.

In contrast, the proposed approach introduces a representation learning stage prior to clustering. The graph is first embedded into a 32-dimensional space using the Node2Vec algorithm, which captures local and global structural patterns through biased random walks. This embedding transforms the graph into a format more suitable for distance-based clustering methods. The embedded nodes are then clustered using the K-Means algorithm.

A fundamental parameter in K-Means clustering is the number of clusters, $k$, which must be specified beforehand. In this study, we selected $k = 17$ based on domain-informed expectations about the spatial complexity of the environment. Although the building's exact number of floors is not used during training—maintaining the unsupervised nature of the method—prior knowledge suggests that similar buildings often consist of 5 to 20 floors. Furthermore, intra-floor separations such as disconnected zones or wings may cause the effective number of spatial groupings to exceed the number of physical floors.

Thus, $k = 17$ was chosen as a reasonable approximation, reflecting both vertical and lateral separability within the environment. Preliminary tests also indicated that this value offered a good trade-off between under- and over-segmentation in the clustering output.

The K-Means algorithm was applied to the Node2Vec embeddings using this $k$ value, with a fixed random seed to ensure reproducibility. The resulting cluster assignments were mapped back to their original trajectory identifiers and grouped accordingly. These results were exported in CSV format for subsequent evaluation.

In parallel, the aforementioned baseline community detection methods were evaluated on the same trajectory graph. Unlike the proposed pipeline, these algorithms operate solely on the original graph without any embedding step. While this avoids assumptions about the geometry of the embedding space or the need to specify $k$, it also limits their ability to capture more subtle structural nuances that the embedding-based approach can reveal.

In summary, the proposed method combines graph construction, feature learning, and clustering to uncover spatial groupings in indoor trajectory data. While it introduces additional parameters and assumptions, such as embedding dimensionality and cluster count, it achieves superior performance in identifying same-floor groupings when compared to classical community detection methods.

\section{Results and Analysis}

We evaluated the clustering performance of several community detection algorithms, including Fast Greedy, Infomap, Label Propagation, Leiden, and Louvain, in comparison to a Node2Vec-based approach followed by k-means clustering (see Table~\ref{tab:clustering_performance}). The evaluation metrics include Adjusted Rand Index (ARI), Normalized Mutual Information (NMI), V-measure, and clustering accuracy after mapping clusters to ground truth labels (i.e., floor assignments). We also considered cluster purity to assess intra-cluster homogeneity.

\vspace{0.5em}
\noindent\textbf{1. Node2Vec Embedding-Based Clustering:}

The Node2Vec-based clustering approach clearly outperformed all community detection algorithms across all metrics. It achieved an ARI of 0.4297 (raw) and 0.5719 (mapped), NMI of 0.5747 (raw) and 0.6131 (mapped), and an overall accuracy of 0.6897. The F1-score reached 0.6199, indicating a balanced performance between precision and recall. Despite a slightly lower average cluster purity (79.50\%) compared to other methods, this approach demonstrated superior consistency with the ground truth, suggesting that it effectively captured both the local and global structure of the graph.

\vspace{0.5em}
\noindent\textbf{2. Community Detection Algorithms:}

All classical community detection methods yielded significantly lower performance. Specifically, ARI scores for these methods were close to or even below zero, indicating near-random or poor agreement with the ground truth:

\begin{itemize}
	\item \textbf{Fast Greedy:} Lowest performance among all methods, with F1-score of 0.3555 and raw ARI of -0.0020. Despite high average purity (94.19\%), its low completeness suggests fragmented clustering with respect to the true classes.
	\item \textbf{Infomap and Label Propagation:} These methods provided slightly better F1-scores (0.4013 and 0.3882, respectively) and modest ARI values (0.0928 and 0.0764), but still far below Node2Vec. Their average cluster purity was 82.45\% and 86.19\%, respectively.
	\item \textbf{Leiden and Louvain:} Nearly identical results, both achieving an F1-score of 0.3651, ARI of 0.0660, and average purity above 92\%. This similarity reflects the shared modularity optimization approach of the two algorithms.
\end{itemize}

Notably, while community detection methods generally produced high cluster purity (ranging from 82\% to 94\%), this did not translate into higher ARI, NMI, or F1-score values. This discrepancy suggests that although some clusters were internally coherent, they failed to align well with the ground truth labels overall. In particular, the homogeneity was consistently higher than completeness, indicating that the algorithms tended to form small pure clusters that did not fully capture all instances of each class.

These results indicate that the Node2Vec + k-means pipeline is more effective than traditional community detection methods for capturing meaningful structural and semantic patterns in the graph. Embedding-based techniques that encode node similarity and graph topology in a continuous space provide a richer representation, which in turn leads to more accurate and interpretable cluster assignments. For applications involving indoor positioning or spatial behavior modeling, such approaches appear to offer substantial advantages.

\begin{table}[htbp]
	\centering
	\caption{Performance comparison of clustering methods}
	\begin{tabular}{lccccc}
		\toprule
		\textbf{Algorithm} & \textbf{Accuracy} & \textbf{F1} & \textbf{ARI} & \textbf{NMI} & \textbf{Purity} \\
		\midrule
		\textbf{Node2Vec} & \textbf{0.6897} & \textbf{0.6199} & \textbf{0.5719} & \textbf{0.5747} & 79.50\% \\
		Fast Greedy & 0.4448 & 0.3555 & 0.0542 & 0.1382 & \textbf{94.19\%} \\
		Infomap & 0.4597 & 0.4013 & 0.0928 & 0.1696 & 82.45\% \\
		Label Prop. & 0.4533 & 0.3882 & 0.0764 & 0.1579 & 86.19\% \\
		Leiden & 0.4437 & 0.3651 & 0.0660 & 0.1309 & 92.37\% \\
		Louvain & 0.4437 & 0.3651 & 0.0660 & 0.1307 & 92.78\% \\
		\bottomrule
	\end{tabular}
	\label{tab:clustering_performance}
\end{table}

\begin{figure}[ht]
	\centerline{\includegraphics[width=1\columnwidth]{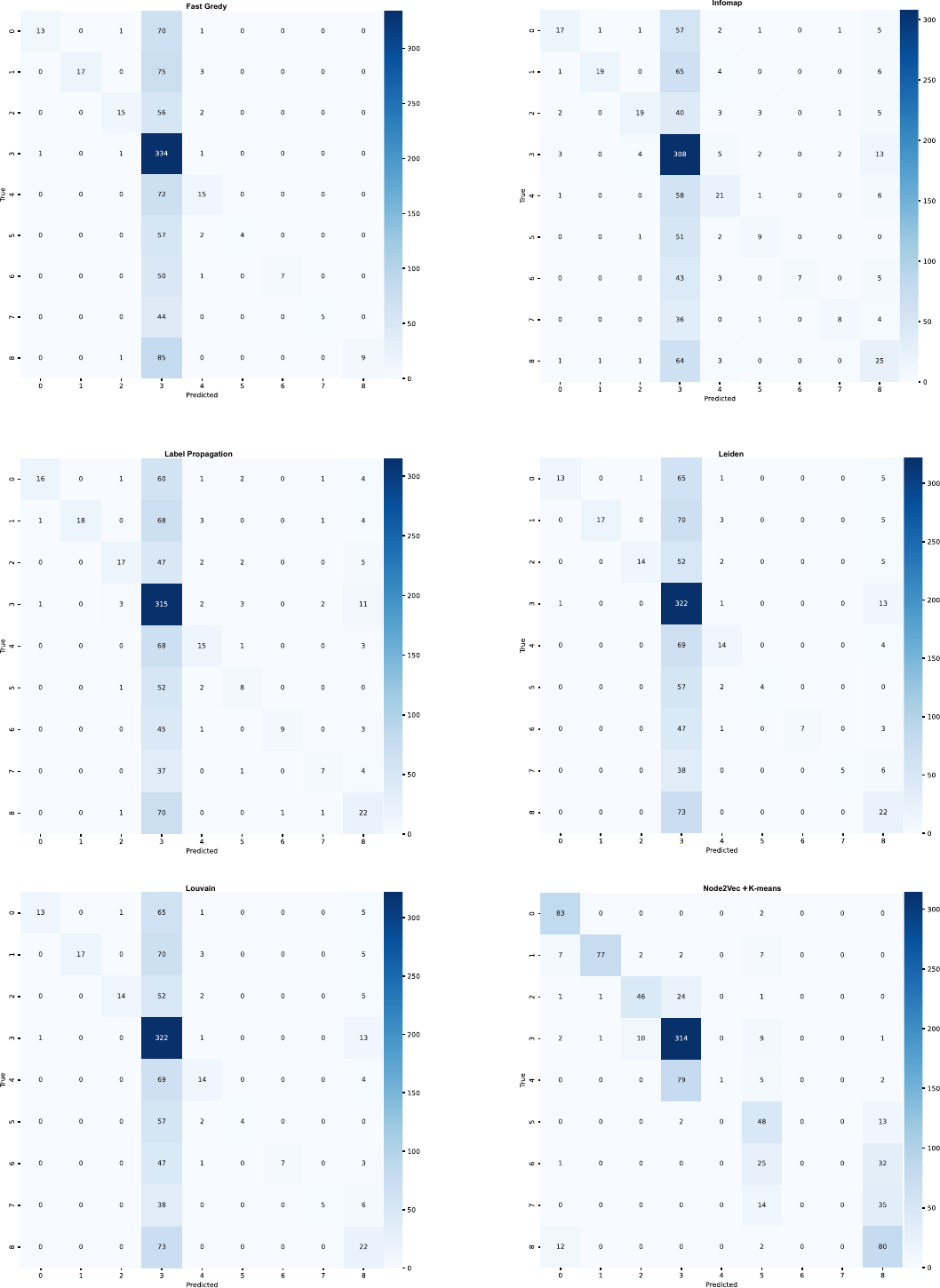}}
	\caption{Confusion matrices of six clustering and community detection methods.}
	\label{fig:confusion_matrices}
\end{figure}

Figure~\ref{fig:confusion_matrices} presents the confusion matrices for six different clustering and community detection methods: Fast Greedy, Infomap, Label Propagation, Leiden, Louvain, and Node2Vec combined with K-means. These matrices allow a direct comparison between the predicted cluster assignments and the ground truth labels.

Among the evaluated methods, the \textbf{Leiden} and \textbf{Louvain} algorithms yielded the most accurate and stable results. Both methods demonstrate strong diagonal dominance across all classes, indicating a high level of agreement with ground truth labels. In particular, the Leiden algorithm exhibits slightly less dispersion across off-diagonal elements, suggesting better cluster purity and less overlap between communities.

The \textbf{Node2Vec + K-means} approach also performed well, especially for the denser clusters such as classes 0, 1, 3, and 4. However, some confusion is observed in the smaller or more sparse classes (e.g., classes 6 and 8), which may result from embedding limitations in low-density regions.

\textbf{Label Propagation} and \textbf{Infomap} showed moderate performance, with increased off-diagonal activity, indicating less precise cluster boundaries. While both algorithms were able to detect the dominant clusters reasonably well (e.g., class 3), they struggled with consistent separation in other categories.

Finally, the \textbf{Fast Greedy} algorithm performed the worst in terms of cluster assignment accuracy. The confusion matrix shows significant dispersion, reflecting high rates of misclassification and weaker alignment with the actual class structure.

In conclusion, Figure~\ref{fig:confusion_matrices} supports that the Leiden and Louvain methods are best suited for the studied graph structure, achieving high accuracy with minimal overlap. The Node2Vec + K-means combination also proves to be a robust alternative, especially when enhanced with embedding techniques. These findings guide future applications in selecting appropriate community detection strategies for indoor movement and signal-based clustering tasks.

Overall, the experimental results reveal a clear trade-off between intra-cluster purity and global alignment with the ground truth. Traditional community detection algorithms, while often producing high-purity clusters, tend to suffer from low completeness and fail to capture the broader semantic structure of the data. In contrast, the Node2Vec-based approach, by leveraging structural embeddings, effectively balances local cohesion and global separation, resulting in superior clustering performance across all evaluation metrics. This highlights the importance of integrating graph representation learning into clustering workflows, especially in complex environments such as indoor positioning, where spatial relationships and signal-based proximities are inherently noisy and non-linear. Future research may explore hybrid approaches that combine the strengths of both embedding-based and modularity-driven methods to further enhance robustness and interpretability.

\section{Conclusion}

This study introduces a graph-based floor separation method that leverages Wi-Fi fingerprint trajectories and Node2Vec embeddings, demonstrating superior performance over traditional community detection algorithms. The proposed approach effectively addresses the challenges of vertical localization in multi-storey buildings, achieving robust clustering despite signal noise and architectural complexities. By combining graph construction, representation learning, and K-means clustering, the method captures both local and global structural patterns, resulting in an accuracy of 68.97\% and an F1-score of 61.99\% on the Huawei University Challenge 2021 dataset. The public release of the dataset and code fosters further research in indoor positioning. Future work should focus on enhancing real-time applicability, optimizing parameter selection, and exploring hybrid sensor integration to improve generalizability across diverse indoor environments.

\bibliographystyle{IEEEtran}
\bibliography{references}
\end{document}